\documentclass[letterpaper, 10 pt, peerreviewca]{ieeeconf}

\IEEEoverridecommandlockouts
\usepackage{graphicx,subfigure}
\usepackage{epstopdf}
\usepackage{tikz}
\usepackage{amsmath}
\usepackage{multirow}
\usepackage{amsfonts}
\usepackage{float}
\usepackage{amssymb}
\usepackage{graphicx}

\newtheorem{lemma}{Lemma}
\newtheorem{proposition}{Proposition}
\newtheorem{corollary}{Corollary}
\newtheorem{definition}{Definition}
\newtheorem{remark}{Remark}
\makeatletter

\newcommand{\Rmnum}[1]{\expandafter\@slowromancap\romannumeral #1@}
\makeatother

\begin{document}
 \title{\LARGE  A Novel Decomposition for Control of DC Circuits and Grid Models with Heterogeneous Energy Sources }
 \author{Shuai Wang \& John Baillieul}
 \maketitle
\let\thefootnote\relax\footnotetext{\noindent\underbar{\hspace{0.8in}}\\
Shuai Wang is  with the Division of Systems Engineering and John Baillieul is with the Departments of Mechanical Engineering; Electrical and Computer Engineering, and the Division of Systems Engineering at Boston University, Boston, MA 02115. Corresponding author is John Baillieul (Email: johnb@bu.edu). \newline The work on power systems and grid topology was supported
by NSF EFRI Grant Number 1038230, and work on large
scale integer optimization was supported by DARPA
Contract HR0011-16-C-0115.}
\begin{abstract}
The way in which electric power depends on the topology of circuits with mixed voltage and current sources is examined. The power flowing in any steady-state DC circuit is shown to depend on a minimal set of key variables called fundamental node voltages and fundamental edge currents. Every steady-state DC circuit can be decomposed into a voltage controlled subcircuit and a current controlled subcircuit. In terms of such a decomposition, the $I^2R$ losses of a mixed source circuit are always the sum of losses on the voltage controlled subcircuit and the current controlled subcircuit. The paper concludes by showing that the total power flowing in a mixed source circuit can be found as critical points of the power expressed in terms of the key voltage and current variables mentioned above. The possible relationship to topology control of electric grid operations is discussed.
\end{abstract}
\section{Introduction}
Because of the fast time constants in changing the system state, transmission  line switching  has been used to reduce losses and improve grid security since the 1980s \cite{Koglin}, \cite{Van}. More recently, with the focus on power markets, a good deal of current research on the operation of smart grids has been focused on the co-optimization of network topology and generation in power system
operation, i.e. reducing the generation cost through changing the topology of the network whenever congestion occurs. 

It has been widely accepted that the line switching problem can be formulated as a mixed integer programming (MIP) problem with some binary variables indicating whether the lines of the network are in or out of service \cite{Chen, Fliscounakis}. 
Formulated in this way, the topology reconfiguration problem is NP-hard. To address this challenge, recent work has been aimed at fast heuristic approaches to line switching. References  \cite{Fisher, Fuller, Soroush} show the effectiveness of co-optimizing the generation and the network topology through simulations on the IEEE 118-bus system and  the WECC 179-bus system.  References \cite{Hedman, Ruiz} demonstrate that topology control can be beneficial even while preserving an N-1 reliable network. The total run time of the
heuristic methods is short enough for practical use for day ahead planning, and with further development, these may provide a useful approach to {\em feasibility correction} in optimal power flow (OPF)  calculation, \cite{Barrows}. Despite the enthusiasm with which the research community has pursued heuristic approaches to topology control, satisfactory grid-scale solutions have remained elusive. 

The present paper examines the loading effect of topology reconfiguration in circuits with mixed voltage and current sources and extends our previous work on purely voltage controlled circuits, \cite{Baillieul}, and purely current controlled circuits, \cite{Wang}. The aim is to provide a possible foundation for heuristic of the kind discussed above. The main results of \cite{Baillieul} and \cite{Wang} show that in a voltage-controlled (current-controlled) circuit (see Def. \ref{def:circuit}), switching on an additional conductive line will always increase (decrease)
congestion (i.e.  increase (decrease) the $I^2R$ losses). 

Mathematically, the DC model of  power flow is  equivalent to a current driven network, where power injections are equivalent to  current sources; power flowing through lines is equivalent to current through edges, etc.  See Table 1. 

\begin{center}
\begin{tabular}{ccccc}   \hline
network &  potential   &  flow  &  impedance  & equation \\[0.07in] \hline\hline \\[-0.05in]
 grid & phase $\theta$  & power $P$ & reactance $X$ & $P=\frac{\theta}{X}$  \\[0.07in]  \hline\\[-0.05in]
circuit&  voltage $V$  & current $I$  & resistance $R$ & $I=\frac{V}{R}$    \\[0.07in]  \hline
\end{tabular}
\end{center}
\begin{center}
{Table 1: The equivalence between a current driven circuit and a transmission grid.}
\end{center}

\begin{figure}[htb]
  \begin{minipage}[t]{0.45\linewidth}\centering
    \includegraphics[width=4.5cm]{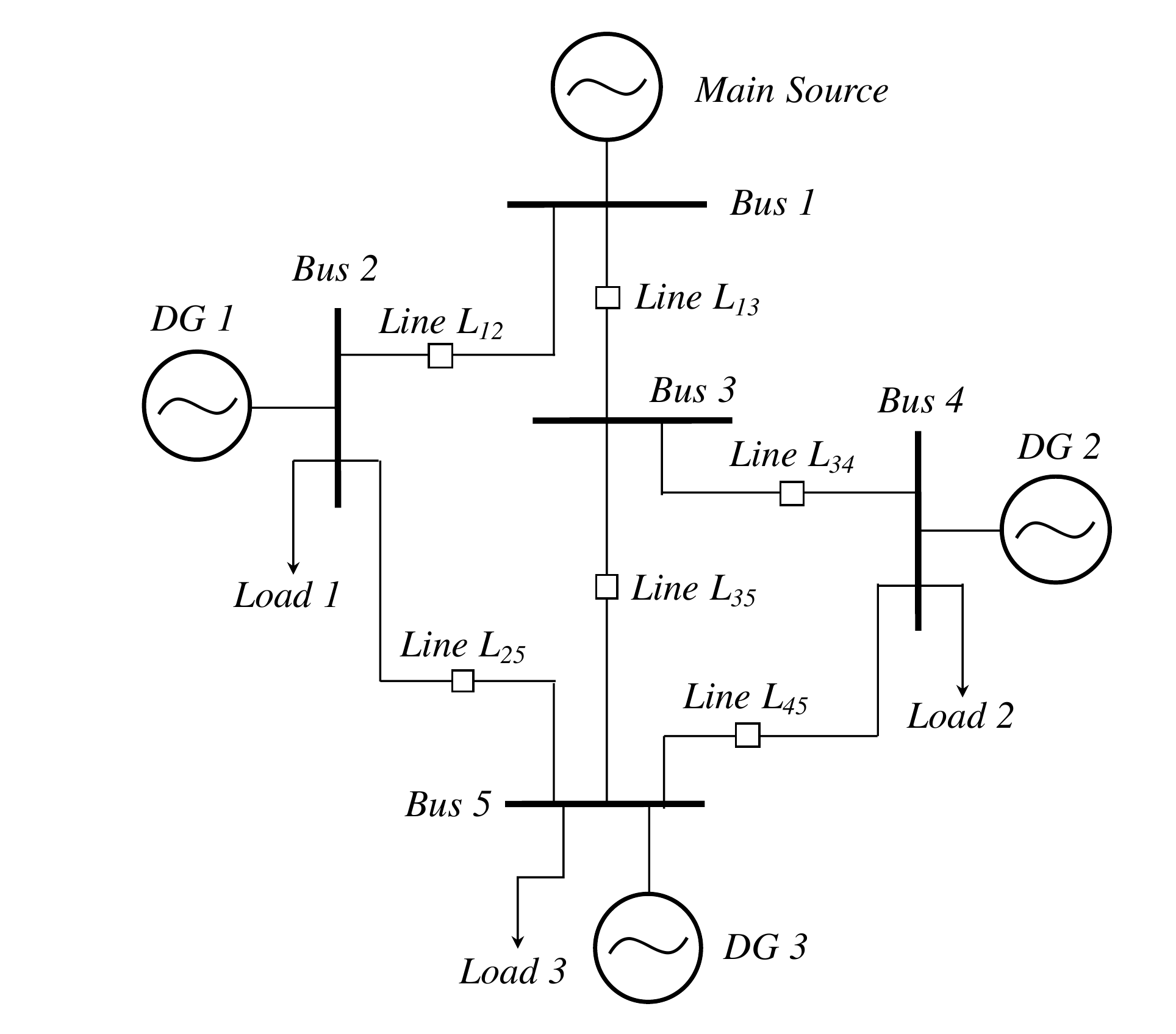}
    \medskip
    \centerline{(a)}
  \end{minipage}\hfill
  \begin{minipage}[t]{0.5\linewidth}\centering
    \includegraphics[width=4cm]{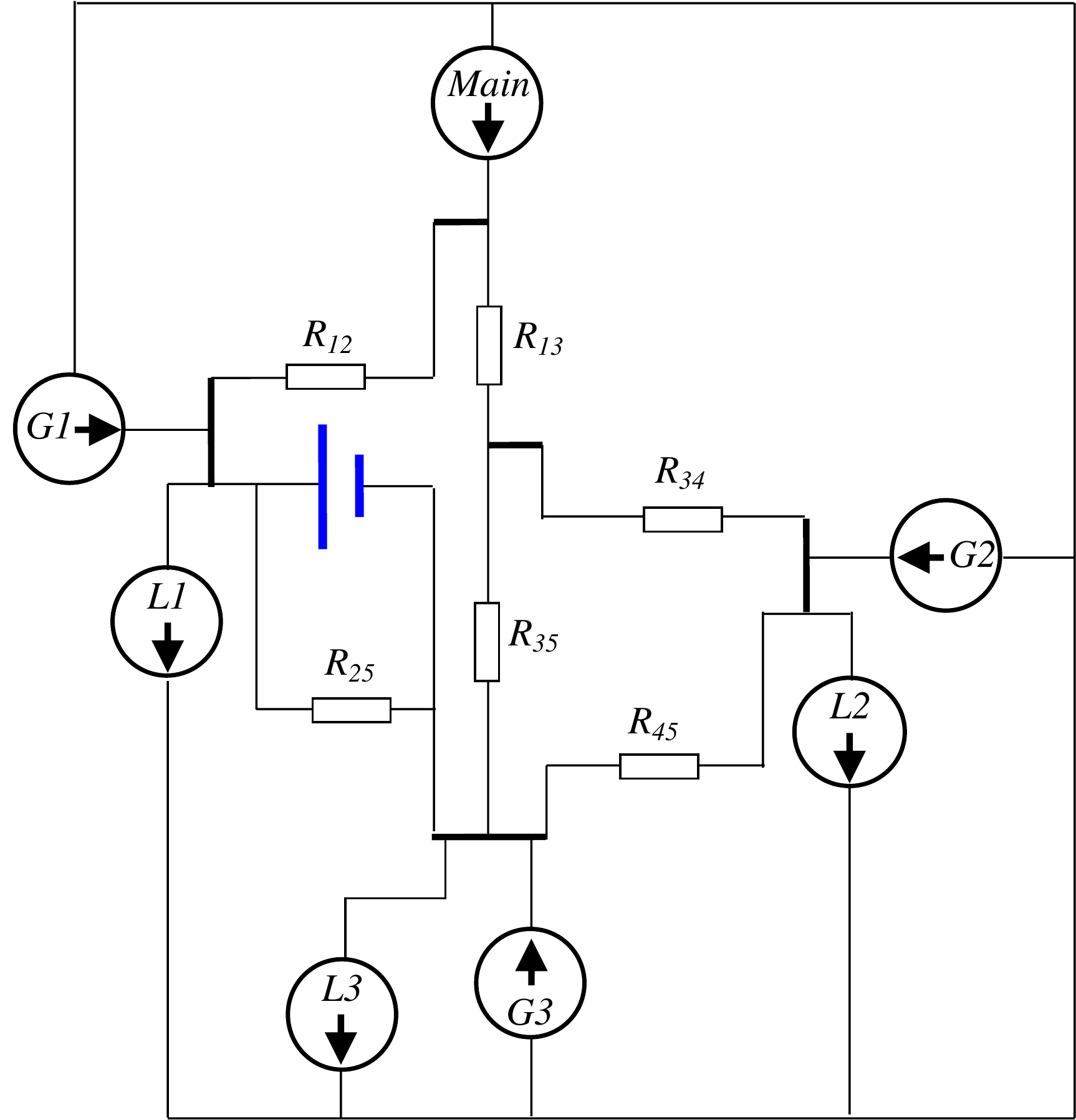}
    \medskip
    \centerline{(b)}
  \end{minipage}
  \caption{(a) A 5-bus network with line overload at $Line$ $L_{25}$. (b) The equivalent mixed-source circuit of (a) with the voltage source (blue) denoting the effect of load regulating equipment at $Bus$ 2 and $Bus$ 5. }
\label{fig:intro}
\end{figure}

While there is a well established correspondence between current-controlled DC-circuits and linearized DC power flow models, the recent increase in load shifting and demand response programs suggest that the formulation of the standard OPF problem should be modified to take advantage of the flexibility  (e.g., loads, reserve requirements, and transmission topology) provided by the smart grid platform. Our mixed source model is better able to capture the features of power grids in which renewables, storage, and demand response
play significant roles. For example, consider the  5-bus  network of Fig. \ref{fig:intro} with a power flow in which there is an overload of $Line$ $L_{25}$. Traditionally, such line overloads can be alleviated either through regulating the generators' output or through dynamic control of the underlying network topology. The increase of demand-side participation and  the development in electrical energy storage in power markets, however, makes it also possible to alleviate the congestion through load regulation or load shifting in time or space or both. In the simplest case, suppose $Bus$ 2 and $Bus$ 5 are equipped with enough energy storage capacities that they are able to release  energy during peak times while storing energy at off-peaks. The effect of such system flexibility can be well abstracted as a ``phase lock" to the overloaded $Line$ $L_{25}$ during peak times which is equivalent to adding a voltage source to the traditional current-controlled circuit model.  
 
Due to the heterogeneity  of the controllers (e.g. the current-control loop and voltage regulator),   distribution networks and microgrids can hardly be modeled as systems with a single class of primary energy sources, pointing out the need for additional research on the mixed source model.

 The present paper looks at whether the topology-dependent loading  phenomena  similar to \cite{Baillieul} and \cite{Wang} can be found  in mixed-source networks.
It is organized as follows.  In the next section (Section
II), we review the needed background on the topology
of DC electric circuits. In particular, we revisit the calculations of total $I^2R$ loss for arbitrary voltage-controlled circuits  and  arbitrary current-controlled circuits, respectively. In Section III, we extend the discussion from single-source circuits (see Def. \ref{def:circuit}) to mixed-sourced circuits. It is shown that the effect of removing an edge from a mixed-source
circuit can be perfectly decomposed into two sub-effects in its voltage-controlled sub-circuit (see Def. \ref{def:mix}) and current-controlled sub-circuit (see Def. \ref{def:mix}), respectively. Meanwhile, it gives a simple method to calculate the change of total $I^2R$ loss for mixed-source circuit based on its reduced equivalent circuit. It is also shown that the total loss of a mixed-source circuit is exactly the sum of total loss of its voltage-controlled sub-circuit and current-controlled sub-circuit. Section IV explores this in terms of four different approaches to calculate the total $I^2R$ loss of an arbitrary mixed-source circuit.  It is shown that all of them are mathematically equivalent, pointing out a  way to convert a certain type of constrained linear programming formulation to an  unconstrained non-linear programming problem. Concluding remarks and possible implications for power networks
are contained in Section V.
\section{Preliminaries}
\begin{definition} A {\em voltage-controlled circuit} is comprised purely by resistors and voltage sources. A {\em current-controlled circuit} is comprised purely by resistors and current sources. Both voltage-controlled circuits and current-controlled circuits are called {\em single-source circuits}.  A circuit that has both current sources and voltage sources is called a {\em mixed-source circuit}.
\label{def:circuit}\end{definition}

\begin{remark}  As capacitors (inductors) act identically as open (short) circuits in DC steady state, the above definition can be easily extended to general steady-state RCL circuits as was shown in \cite{Baillieul}  and \cite{Wang}.
\end{remark}

\begin{definition}  An {\em edge} in a network graph represents a single element either a voltage source, a current source or a resistor.
A {\em node} denotes the position of connection where two or more
edges meet. A {\em cycle} is
any closed path. \end{definition}

\begin{figure}[htb]
  \begin{minipage}[t]{0.3 \linewidth}\centering
    \includegraphics[width=1.6cm]{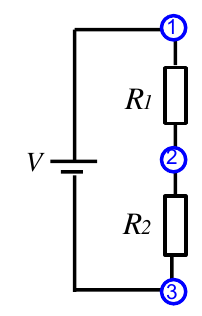}
    \centerline{(a)}
  \end{minipage} \hspace{.1in}
  \begin{minipage}[t]{0.3\linewidth}\centering
    \includegraphics[width=2cm]{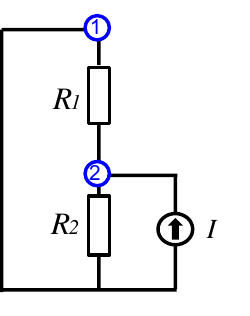}
    \centerline{(b)}
  \end{minipage}
    \begin{minipage}[t]{0.33\linewidth}\centering
    \includegraphics[width=2.6cm]{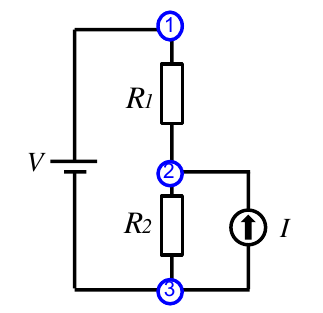}
    \centerline{(c)}
  \end{minipage}
  \caption{(a) A voltage-controlled circuit with nodes $\{1,2,3\}$. (b)  A current-controlled circuit with nodes $\{1,2\}$. (c)  A mixed-source circuit with  nodes $\{1,2,3\}$.} 
\label{fig:example}
\end{figure}

\subsection{Voltage-Controlled Circuit \cite{Baillieul}}
\begin{definition}\cite{Baillieul}
For a voltage controlled circuit, a {\em fundamental node basis} is a maximal set of nodes among which there exist  no paths comprised purely of voltage source edges. Their voltages are called {\em fundamental nodal voltages}.
\end{definition}

\begin{figure}[htbp]
	\centering
		\includegraphics[width=0.2 \textwidth]{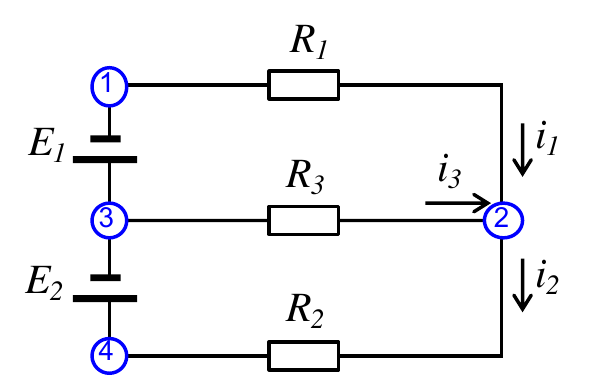}
	\caption{ A simple circuit with DC-voltage sources and resistive loads.} \label{keyfigure}
\end{figure}

The fundamental node basis may not be unique for a voltage-controlled circuit, but its dimension is always uniquely determined.  For example, the fundamental node basis for the circuit in Fig. \ref{keyfigure}  can be nodes  $\{1,2\}$,  $\{3,2\}$ or $\{4,2\}$.

By definition, if one node, say the $i$-th node, is not included in the selected fundamental node basis, there must exist one and only one fundamental node that has a pure voltage source path connecting it and the $i$-th node. Thus, once all fundamental nodal voltages are known, all other nodal voltages can be determined by adding the voltage contributions from the voltage source edges that connect them to the fundamental nodes.

There may or may not exist a pure voltage source path between the pair of endpoints of a resistor edge. If such a path exists, then the voltage drop between its endpoints is always fixed. We call such a resistor a {\em Type V1} resistor. We can use a scalar, say $P_{V1}$, to denote the total $I^2R$ loss of such resistor edges.

For a resistor edge that doesn't have a pure voltage source path between its  endpoints, its $I^2R$ loss may vary if we change the connectivity of another edge. We call such a resistor a {\em Type V2} resistor. We assume, without   loss of generality, that the endpoints of each Type V2 resistor are connected  to a pair of fundamental nodes, $\{i,j\}$ ($i,j=1,...,D$, where $D$ is the cardinality of the fundamental node basis), and these are connected by a voltage source path. In general, there may be more than one Type V2 resistor edge  connected to fundamental node pair $\{i,j\}$, and we denote the total number of such resistor edges by $L_{i,j}$. Here, we assume the $k$-th resistor edge discussed above has value  $R_{i,j,k}$   $(k=1,...,L_{i,j}$).

It is easy to see that the sets of Type V1  resistors and Type V2  resistors are jointly exhaustive. A potential function denoting the total $I^2R$ loss of all resistors on a voltage controlled circuit can then be formed by:
\begin{equation}
 P_V =P_{V1}+\sum_{i=1}^{D-1}\sum_{j=i+1}^{D}\sum_{k=1}^{L_{i,j}}\frac{(e_{v_i}+e_{P_{v_{i,k}}}-e_{v_j}-e_{P_{v_{j,k}}})^2}{R_{i,j,k}}
 \label{loss:voltage}
\end{equation}
where $\{e_{v_1},...,e_{v_{M}}\}$ are the fundamental nodal voltages, and $e_{P_{v_{i,k}}}$ denotes the  algebraic sum of voltages on the pure voltage source path connecting the fundamental node $i$ and one endpoint of the resistor $R_{i,j,k}$. Similarly, $e_{P_{v_{j,k}}}$  is the sum of voltages along the path connecting the other endpoint of the resistor to the fundamental node $j$.

This can be best understood by using the circuit in Fig. \ref{keyfigure}  as an illustrative example. Clearly, there are no resistor edges in  Fig. \ref{keyfigure} with endpoints being  connected by a pure voltage source path, i.e. $P_{V1}=0$. The dimension of the fundamental node basis has been shown to be 2, i.e $D=2$, and we randomly choose nodes $\{1,2\}$ as the fundamental node basis. The endpoints of three resistor edges $\{R_1,R_2,R_3\}$  are either directly connected to node 1 and node 2 or  indirectly connected to node 1 by a pure voltage source path, i.e. $L_{1,2}=3$, $R_{1,2,1}=R_1$, $R_{1,2,2}=R_2$, and $R_{1,2,3}=R_3$. For example, the left endpoint of $R_2$ (namely $R_{1,2,2}$) is connected by a path of $E_1$ and $E_2$ to node 1, i.e.  $e_{P_{v_{1,2}}}=-E_1-E_2$, and its right endpoint is directly connected to node 2, i.e.  $e_{P_{v_{2,2}}}=0$. Denoting the voltages at node 1 and 2 as $e_{v1}$ and  $e_{v2}$, we know the $I^2R$ loss of $R_2$ is  given by $\frac{(e_{v_1}+e_{P_{v_{1,2}}}-e_{v_2}-e_{P_{v_{2,2}}})^2}{R_{1,2,2}}$. Repeating the calculation for all three resistors, we   have 
\[
\begin{split}
  P_V =\sum_{k=1}^{3}\frac{(e_{v_1}+e_{P_{v_{1,k}}}-e_{v_2}-e_{P_{v_{2,k}}})^2}{R_{1,2,k}}
 \end{split}
\]
\subsection{Current-Controlled Circuit \cite{Wang}}

\begin{definition}
\cite{Liebchen} If there exists some spanning tree $T$ for a given graph, and $e$ denotes an edge that is not in $T$, then the simple cycle consisting of $e$ together with the path in $T$ connecting the endpoints of $e$ is called the {\em fundamental cycle} defined by $e$. A cycle basis formed in this way is called a {\em fundamental cycle basis}. 
\end{definition}

\begin{definition} Given an arbitrary DC circuit network consisting of current sources, voltage sources and resistors, its {\em resistance graph} is formed by:
\begin{itemize}
\item replacing the original position of every current source
with its internal resistance: an open circuit;
\item replacing the original position of every voltage source
with its internal admittance: a  short circuit.
\end{itemize}
\end{definition}

For a current-controlled circuit, its resistance graph can be   formed by removing all source edges. A spanning tree ${\cal T}$  of its resistance graph can be   found by using depth-first search.  Then a fundamental cycle basis of the resistance graph can be formed based on ${\cal T}$. We assume without loss of generality that there are $N$ fundamental cycles, and the edges that define the fundamental  cycles are $\{e_1,e_2,...,e_N\}$. The original current-controlled circuit  can be reconstructed from the resistance graph by putting back all current source edges. With all current sources operating, we denote the current flowing on the edge $e_i$ that defines the $i$-th fundamental cycle  by $I_{e_i}$. 

A resistor edge, if one of its endpoints is a leaf vertex, doesn't belong to any fundamental cycle of the resistance graph. In such case, its $I^2R$ loss  will always be unchanged as the current flowing through the resistor is fixed. We call such resistors {\em Type }I$1$ resistors. We can use a scalar, say $P_{I1}$, to denote the total loss of such resistors.

A resistor edge, if neither of its endpoints is a leaf vertex, may belong to either one or several fundamental cycles of the resistance graph. For a resistor edge that is exclusively owned by one fundamental cycle, say the $i$-th fundamental cycle, we call it a {\em Type I2} resistor. We can denote the number of the $i$-th fundamental cycle's exclusive edges by $O_{i}$ ($i=1,...,N$), and  the resistance of the $k$-th exclusive edge as $R_{i,i,k}$, then the total loss of such resistor edges can be computed by:
\[
P_{I2} =  \sum_{i=1}^{N}\sum_{k=1}^{O_{i}}(I_{e_i}+I_{P_{e_i,k}})^2R_{i,i,k}
\]
where $I_{P_{e_i,k}}$ denotes the  algebraic sum of current injections from current source edges and/or the Type I1 resistor edges to the path connecting $e_i$ and the $k$-th exclusive edge.

For those resistor edges that are shared by two or more fundamental cycles, we call them {\em Type I3} resistors.  We can denote the number of such resistor edges by $M$,  the resistance of the $k$-th ($k=1,...,M$) edge by $R_{k}$, the number of fundamental cycles that are associated with the $k$-th edge by $n_k$, and the edges defining these associated fundamental cycles by $\{e_{k_1},e_{k_2},...,e_{k_{n_{k}}}\}$ $(1<{k_1},...,k_{n_{k}}<N)$. Then the total loss of such resistor edges can be computed by
\[
P_{I3}
=\sum_{k=1}^{M}(\sum_{i=1}^{n_k} I_{e_{k_{i}}}+I_{P_k})^2R_{k}
\]
where $I_{P_k}$ denotes the  algebraic sum of current injections from current source edges and/or the Type I1 resistor edges to the paths connecting the edges defining the associated fundamental cycles and the $k$-th Type I3 edge.

It is easy to see that the sets of Type I1  resistors, Type I2  resistors and Type I3  resistors are jointly exhaustive. Then a potential function denoting the total loss of all resistors in a current-controlled circuit can be given by:
\begin{equation}
P_I
=P_{I1}+P_{I2}+P_{I3}\label{loss:current}
\end{equation}

\begin{definition}  \cite{Wang} The adding (removing) of an edge to (from) an existing graph is called a parallel attachment (removal) if the node set of the graph is unchanged but the number of fundamental cycles is increased (decreased) by 1 after the operation. The adding (removing) of an edge to (from) an existing graph is called a serial attachment (removal) if the  cycle space of the graph is unchanged but the number of nodes is increased by one after the operation.   \end{definition}

\begin{figure}[htbp]
	\centering
		\includegraphics[width=0.45\textwidth]{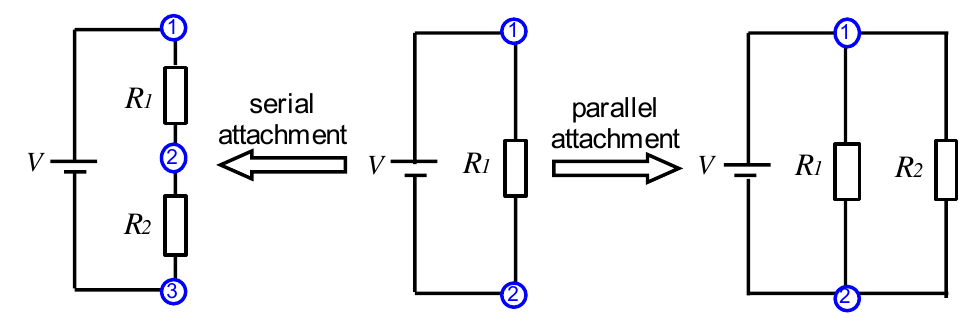}
	\caption{An example of parallel attachment and serial attachment.} \label{serialparallel}
\end{figure}

In graph theory, a serial attachment can be viewed as the subdivision of some edge. Fig. \ref{serialparallel} shows an example of  parallel attachment and  serial attachment of a resistor $R_2$ to a circuit comprised by a voltage source $V$ and a resistor $R_1$.

\begin{definition} \cite{Wang}  An electric element in the circuit is called a
{\em passive} element if its current and voltage are of opposite
polarity (and therefore the element consumes power), and an
{\em active} element if its current and voltage are of same polarity
(and therefore the element delivers power). \end{definition}

\section{The Case of Mixed-Source Networks}
\begin{figure}[htb]
  \begin{minipage}[t]{0.49\linewidth}\centering
    \includegraphics[width=4cm]{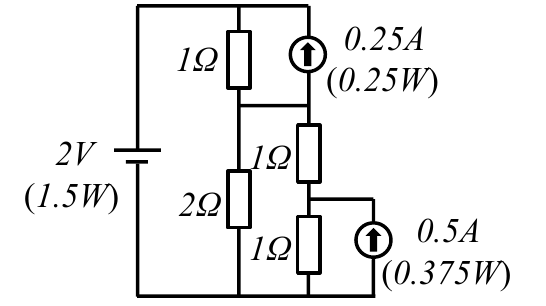}
    \medskip
    \centerline{(a)}
  \end{minipage}\hfill
  \begin{minipage}[t]{0.49\linewidth}\centering
    \includegraphics[width=4cm]{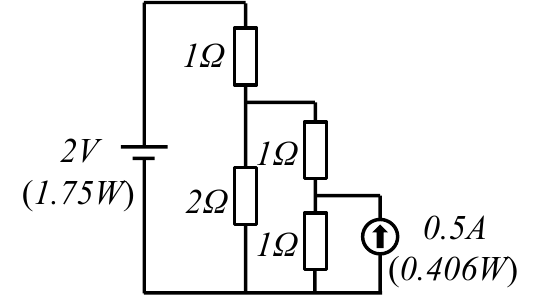}
    \medskip
    \centerline{(b)}
  \end{minipage}
  \caption{(a) A mixed-source network with two current sources and one voltage source. (b)  A mixed-source network with one current source  and one voltage sources. (Example from \cite{Wang}).} 
\label{fig:paradox}
\end{figure}

In \cite{Baillieul} and \cite{Wang}, we showed that the parallel attachment of an active current (voltage) source always increases the total loss of a current-controlled (voltage-controlled) circuit. It is natural to ask if similar results can be generalized for an arbitrary mixed-source circuit. The answer is no. \cite{Wang} shows a simple counterexample which is revisited in Fig. \ref{fig:paradox}. An interesting paradoxical behavior happens in Fig. \ref{fig:paradox}: the removal of an active current source causes a redistribution of the current that results in higher total $I^2R$ loss of a mixed-source circuit.
 
In order to explore this paradox, we have the following:

\begin{definition}
A {\em source factor} in a circuit is defined to be the sensitivity of the current flowing through a voltage source (or of voltage difference between the endpoints of a current source) with respect to a change in the value of another voltage source or current source.\label{def:coefficient}
\end{definition}

By above definition, we have the following notations:
\begin{itemize}
\item the sensitivity of the current flowing through voltage source $i$ with respect to a change in the value of voltage source $j$  is denoted as $s^{\mathbb{V}\mathbb{V}}_{j,i}$;
\item the sensitivity of the current flowing through  voltage source $i$  with respect to a change in the value of current source $j$   is denoted as $s^{\mathbb{I}\mathbb{V}}_{j,i}$;
\item the sensitivity of the voltage difference between the endpoints of current source $i$  with respect to a change in the value of  voltage source $j$  is denoted as $s^{\mathbb{V}\mathbb{I}}_{j,i}$;
\item the sensitivity of the voltage difference between the endpoints of  current source $i$  with respect to a change in the value of  current source $j$   is denoted as $s^{\mathbb{I}\mathbb{I}}_{j,i}$.
\end{itemize}

\begin{figure}[htbp]
	\centering
		\includegraphics[width=0.3\textwidth]{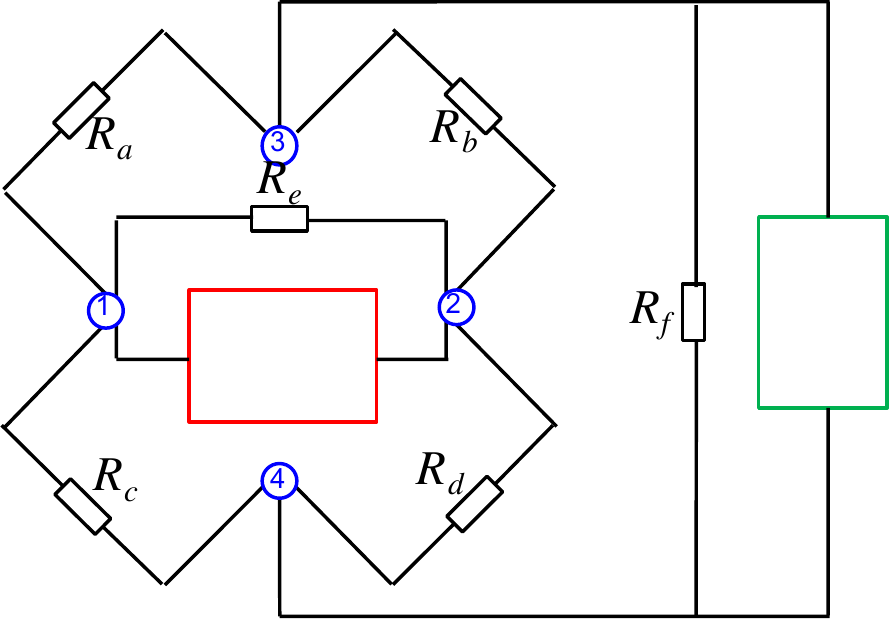}
	\caption{The equivalent reistance network for an arbitrary 4-terminal resistance network.} \label{4terminal}
\end{figure}

By the Generalized Notorn Theorem \cite{Hosoya},  any 4-terminal resistance graph can be reduced to an equivalent resistance network. The  equivalent network  consists of 6 resistors and its graph is the complete one as shown in Fig. \ref{4terminal}. 

The red box (and green box) in Fig. \ref{4terminal} can then be filled with  a voltage source or current source in order to study some useful properties of the source factor. Basically, we have following results (whose proof is a basic calculation based on Fig. \ref{4terminal} and thus is omitted here):
\begin{equation}
\begin{split}
 &s^{\mathbb{V}\mathbb{V}}_{j,i} =  s^{\mathbb{V}\mathbb{V}}_{i,j} \\&
  s^{\mathbb{I}\mathbb{V}}_{j,i}  =  -s^{\mathbb{V}\mathbb{I}}_{i,j} \\&
 s^{\mathbb{I}\mathbb{I}}_{j,i}  =  s^{\mathbb{I}\mathbb{I}}_{i,j} .
 \end{split}\label{equ:factor}
\end{equation}

\begin{lemma}
The parallel attachment of a resistor edge to endpoint pair $\{m,n\}$ in a mixed-source circuit will always decrese the voltage difference between $\{m,n\}$, and will keep the voltage polarity of $\{m,n\}$ unchanged.   \label{lemma:polarity}
\end{lemma} 
\textit{Proof:}  Thevenin's theorem states that any linear circuit with voltage and current sources and  resistances can be replaced at terminals $m$-$n$ by an equivalent voltage source $V_{mn}$ in series connection with an equivalent resistor $R_{mn}$. $V_{mn}$ is the voltage obtained at terminals $m$-$n$ before we add the new resistor edge. Denoting the new voltage obtained at terminals $m$-$n$ after we add the new resistor edge as $V_{mn}^{'}$, it is easy to prove that   $V_{mn}^{'}$ must be smaller than $V_{mn}$, and the voltage polarity of $\{m,n\}$ must be unchanged.
$\hfill{} \Box$
\begin{definition}
 For a given mixed-source circuit, $C_M$, its \emph{voltage-controlled sub-circuit}, $C_V$, is created by replacing all current source edges with open circuits in $C_M$; and  its \emph{current-controlled sub-circuit}, $C_I$, is created by replacing all voltage source edges with short circuits in $C_M$. It is easy to see that $C_M$, $C_V$, and $C_I$ have the same set of resistance edges. To prevent confusion, we denote the $i$-th resistance edge by $R_i^M$ in $C_M$, $R_i^V$ in $C_V$, and $R_i^I$ in $C_I$, respectively, and denote the current flowing on  the $i$-th resistance edge by $I_i^M$ in $C_M$, $I_i^V$ in $C_V$, and $I_i^I$ in $C_I$, respectively.\label{def:mix}
\end{definition}

For example, the circuits in Fig. \ref{fig:example}(a) and  Fig. \ref{fig:example}(b) are  the voltage-controlled sub-circuit and current-controlled sub-circuit of the circuit  in Fig. \ref{fig:example}(c), respectively.
\begin{proposition}
The change of total losses, $\Delta P$, resulting from the parallel removal (parallel attachment) of a resistance link $R_j^M$ from a mixed-source circuit is given by $\Delta P=\Delta P_V + \Delta P_I$, where $\Delta P_V$ denotes the change of losses resulting from removing (adding) the link $R_j^V$ from its voltage-controlled sub-circuit, and $\Delta P_I$ denotes the change of losses resulting from removing  (adding) the link  $R_j^I$ from its current-controlled sub-circuit.\label{prop:mix2}
\end{proposition} 
\textit{Proof:} It is easy to prove that the parallel attachment and parallel removal have exactly the opposite effect on the total loss of a circuit. The parallel removal part of the proposition   is thus logically equivalent to  the parallel attachment part of the proposition. Hence we just need to prove the parallel removal part of the proposition.

We assume without loss of generality that there are $k$ current sources $\{\mathbb{I}_1,...,\mathbb{I}_k\}$, and $l$ voltage sources $\{\mathbb{V}_1,...,\mathbb{V}_l\}$, in the circuit. Suppose we are going to remove the $j$-th resistor edge $R_j$, and its endpoint pair is $\{m,n\}$. 

By the principle of energy conservation, the change of total $I^2R$ loss must be equivalent to the change of total sources' energy output  
\[\sum_{i=1}^{k}\mathbb{I}_i\Delta V_i+\sum_{i=1}^{l}\mathbb{V}_i\Delta I_i
\]
where $\Delta V_i$ denotes the change of voltage difference between the endpoints of the $i$-th current source, and $\Delta I_i$ denotes the change of current flowing through   the $i$-th voltage source. Here, in order to calculate $\Delta V_i$ and $\Delta I_i$, we replace the  $j$-th resistor edge by a passive current source with value $I_{mn}$, i.e. the current flowing on the resistor edge before its removal.  Clearly, such a replacement increases the number of  source edges by 1 and deceases the number of resistor edges by 1, but it has no effect on the rest of the circuit. The  voltage difference between the endpoints of the new current source edge must be equivalent to $V_{mn}$, i.e. the the voltage difference between node pair $\{m,n\}$ before the removal.

By the superposition principle, it is easy to prove that $V_{mn}$ is a linear combination of $\{\mathbb{I}_1,...,\mathbb{I}_k, I_{mn}\}$ and $\{\mathbb{V}_1,...,\mathbb{V}_l\}$, i.e.
\[
\begin{split}
V_{mn}&=
\begin{bmatrix}
 \mathbb{I}_1  & \cdots  & \mathbb{I}_k & I_{mn} 
\end{bmatrix}\begin{bmatrix}
s^{\mathbb{I}\mathbb{I}}_{1,mn}\\ 
\vdots \\ 
 s^{\mathbb{I}\mathbb{I}}_{k,mn}\\ 
 s^{\mathbb{I}\mathbb{I}}_{mn,mn} 
\end{bmatrix} 
+\begin{bmatrix}
 \mathbb{V}_1  & \cdots  & \mathbb{V}_l  
\end{bmatrix}\begin{bmatrix}
s^{\mathbb{V}\mathbb{I}}_{1,mn}\\ 
\vdots \\ 
 s^{\mathbb{V}\mathbb{I}}_{l,mn}\end{bmatrix}
\end{split}
\]
where $\{s^{\mathbb{I}\mathbb{I}}_{1,mn} , \cdots , s^{\mathbb{I}\mathbb{I}}_{k,mn}, s^{\mathbb{I}\mathbb{I}}_{mn,mn}\} $ and $\{s^{\mathbb{V}\mathbb{I}}_{1,mn} , \cdots , s^{\mathbb{V}\mathbb{I}}_{l,mn}\} $ are source factors. To be more specific, $s^{\mathbb{I}\mathbb{I}}_{u,w}$ ($u,w=1,...,k,mn)$ is   the  sensitivity of  the voltage difference between the $w$-th current source's endpoints with respect to a change in the value of the $u$-th current source.
 $s^{\mathbb{V}\mathbb{I}}_{u,w}$ ($u=1,...,l$ and $ w=1,...,k,mn)$ is   the  sensitivity of  the voltage difference between the $w$-th current source's endpoints with respect to a change in the value of the $u$-th voltage source. Similarly, we have  
\[
\begin{split}
V_i
=&\begin{bmatrix}
 \mathbb{I}_1  & \cdots  & \mathbb{I}_k & I_{mn} 
\end{bmatrix}\begin{bmatrix}
s^{\mathbb{I}\mathbb{I}}_{1,i}\\ 
\vdots \\ 
 s^{\mathbb{I}\mathbb{I}}_{k,i}\\ 
 s^{\mathbb{I}\mathbb{I}}_{mn,i} 
\end{bmatrix}
  +\begin{bmatrix}
 \mathbb{V}_1  & \cdots  & \mathbb{V}_l  
\end{bmatrix}\begin{bmatrix}
s^{\mathbb{V}\mathbb{I}}_{1,i}\\ 
\vdots \\ 
 s^{\mathbb{V}\mathbb{I}}_{l,i}\end{bmatrix}
\end{split}
\]
where $V_i$ ($i=1,...,k$)denotes the voltage difference between the endpoints of the $i$-th current source, and
\[
\begin{split}
I_i
=&\begin{bmatrix}
 \mathbb{I}_1  & \cdots  & \mathbb{I}_k & I_{mn} 
\end{bmatrix}\begin{bmatrix}
s^{\mathbb{I}\mathbb{V}}_{1,i}\\ 
\vdots \\ 
 s^{\mathbb{I}\mathbb{V}}_{k,i}\\ 
 s^{\mathbb{I}\mathbb{V}}_{mn,i} 
\end{bmatrix}
  +\begin{bmatrix}
 \mathbb{V}_1  & \cdots  & \mathbb{V}_l  
\end{bmatrix}\begin{bmatrix}
s^{\mathbb{V}\mathbb{V}}_{1,i}\\ 
\vdots \\ 
 s^{\mathbb{V}\mathbb{V}}_{l,i}\end{bmatrix}
\end{split}
\]
where $I_i$ ($i=1,...,l$)denotes the current flowing through the $i$-th voltage source.

Clearly, the removal of the the $j$-th resistor edge $R_j$ and the removal of the current source $I_{mn}$ have exactly the same effect on the total energy output of the   $k$ current sources and $l$ voltage sources. In addition, the removal of the current source $I_{mn}$ doesn't further change the resistance graph. Thus, we have 
\[
\begin{bmatrix}
\Delta V_1\\ 
\vdots \\ 
\Delta V_k
\end{bmatrix}=
-I_{mn}\begin{bmatrix}
s^{\mathbb{I}\mathbb{I}}_{mn,1}\\ 
\vdots \\ 
s^{\mathbb{I}\mathbb{I}}_{mn,k}
\end{bmatrix}\]\[
\begin{bmatrix}
\Delta I_1\\ 
\vdots \\ 
\Delta I_l
\end{bmatrix}=
-I_{mn}\begin{bmatrix}
s^{\mathbb{I}\mathbb{V}}_{mn,1}\\ 
\vdots \\ 
s^{\mathbb{I}\mathbb{V}}_{mn,l}
\end{bmatrix}
\] 
\[
\begin{split}
&\sum_{i=1}^{k}\mathbb{I}_i\Delta V_i+\sum_{i=1}^{l}\mathbb{V}_i\Delta I_i= 
-I_{mn}\{\begin{bmatrix}
 \mathbb{I}_1  \ \cdots  \ \mathbb{I}_k
\end{bmatrix} 
 \begin{bmatrix}
s^{\mathbb{I}\mathbb{I}}_{mn,1}\\ 
\vdots \\ 
s^{\mathbb{I}\mathbb{I}}_{mn,k}
\end{bmatrix}
+ \begin{bmatrix}
 \mathbb{V}_1  \ \cdots  \ \mathbb{V}_l
\end{bmatrix}
 \begin{bmatrix}
s^{\mathbb{I}\mathbb{V}}_{mn,1}\\ 
\vdots \\ 
s^{\mathbb{I}\mathbb{V}}_{mn,l}
\end{bmatrix}\}.
\end{split}
\]

By (\ref{equ:factor}), we have  $s^{\mathbb{I}\mathbb{I}}_{mn,i}=s^{\mathbb{I}\mathbb{I}}_{i,mn}$ ($i=1,...,k,mn$) and   $s^{\mathbb{I}\mathbb{V}}_{mn,i}=-s^{\mathbb{V}\mathbb{I}}_{i,mn}$ ($i=1,...,l$)
which further gives us  
\[
\begin{split}
&\sum_{i=1}^{k}\mathbb{I}_i\Delta V_i+\sum_{i=1}^{l}\mathbb{V}_i\Delta I_i=
-I_{mn}\{\begin{bmatrix}
 \mathbb{I}_1  \ \cdots \ \mathbb{I}_k
\end{bmatrix} 
 \begin{bmatrix}
s^{\mathbb{I}\mathbb{I}}_{1,mn}\\ 
\vdots \\ 
s^{\mathbb{I}\mathbb{I}}_{k,mn}
\end{bmatrix}
- \begin{bmatrix}
 \mathbb{V}_1  \ \cdots  \ \mathbb{V}_l
\end{bmatrix}
 \begin{bmatrix}
s^{\mathbb{V}\mathbb{I}}_{1,mn}\\ 
\vdots \\ 
s^{\mathbb{V}\mathbb{I}}_{l,mn}
\end{bmatrix}\}.
\end{split}
\]

By replacing all voltage source edges with short circuits (i.e. setting the value of all voltage sources to zero), we get the current-controlled sub-circuit, $C_I$. It is easy to prove that  $\Delta P_I$, the change of $I^2R$ loss  resulting from removing  $R_j$ from the current-controlled sub-circuit, is given by 
\[
\Delta P_I=
-I_{mn}\begin{bmatrix}
 \mathbb{I}_1  & \cdots  & \mathbb{I}_k
\end{bmatrix} 
 \begin{bmatrix}
s^{\mathbb{I}\mathbb{I}}_{1,mn}\\ 
\vdots \\ 
s^{\mathbb{I}\mathbb{I}}_{k,mn}
\end{bmatrix}.
\]  
Similarly, we have
\[
\Delta P_V=
I_{mn} \begin{bmatrix}
 \mathbb{V}_1  & \cdots  & \mathbb{V}_l
\end{bmatrix}
 \begin{bmatrix}
s^{\mathbb{V}\mathbb{I}}_{1,mn}\\ 
\vdots \\ 
s^{\mathbb{V}\mathbb{I}}_{l,mn}
\end{bmatrix}.
\]
This ends the proof. $\hfill{} \Box$

\begin{proposition}
The change of total losses, $\Delta P$, resulting from the serial removal (serial attachment) of a resistance link $R_j^C$ from a mixed-source circuit is given by $\Delta P=\Delta P_V + \Delta P_I$, where $\Delta P_V$ denotes the change of losses resulting from removing (adding) the link $R_j^V$ from its voltage-controlled sub-circuit, and $\Delta P_I$ denotes the change of losses resulting from removing (adding) the link  $R_j^I$ from its current-controlled sub-circuit.\label{prop:serial}
\end{proposition} 
\textit{Proof:} As in Proposition \ref{prop:mix2}, the serial removal part of the proposition   is   logically equivalent to  the serial attachment part of the proposition. So we just need to prove the serial removal part of the proposition. 

Again, we assume  there are $k$ current sources $\{\mathbb{I}_1,...,\mathbb{I}_k\}$, and $l$ voltage sources $\{\mathbb{V}_1,...,\mathbb{V}_l\}$ in the circuit. Suppose we are going to remove the $j$-th resistor edge $R_j$, and its endpoint pair is $\{m,n\}$.   Since we are doing a  serial removal, it will merge node $m$ and node $n$ together. Electrically, it is also equivalent to  the parallel attachment of a zero resistance edge to $\{m,n\}$. Proposition \ref{prop:mix2} states that the change of total $I^2R$ loss resulting from the parallel attachment of a zero resistance edge to $\{m,n\}$ is given by $ \Delta P_V + \Delta P_I$.  
$\hfill{} \Box$

We now state a sequence of corollaries that follow fairly directly from Proposition \ref{prop:mix2} and  \ref{prop:serial}. 
\begin{corollary}
The change of total $I^2R$ loss, $\Delta P_V$, resulting from the parallel removal of a resistor edge with endpoint pair $\{m,n\}$ from a voltage-controlled circuit is given by $\Delta P_V=I_{mn}V_{mn}^{'}$, where $I_{mn}$ denotes the current flowing on the edge before its removal, and $V_{mn}^{'}$ denotes the voltage difference between node pair $\{m,n\}$ after its removal. \label{prop:voltage}
\end{corollary} 
\textit{Proof:} At the end of the proof for Proposition \ref{prop:mix2}, we know that  $\Delta P_V$, the change of $I^2R$ loss  resulting from removing  $R_j$ from the voltage-controlled sub-circuit, is given by 
\[
\Delta P_V=
I_{mn} \begin{bmatrix}
 \mathbb{V}_1  & \cdots  & \mathbb{V}_l
\end{bmatrix}
 \begin{bmatrix}
s^{\mathbb{V}\mathbb{I}}_{1,mn}\\ 
\vdots \\ 
s^{\mathbb{V}\mathbb{I}}_{l,mn}
\end{bmatrix}.
\]  
Since there are no current sources  in a voltage-controlled circuit, we have
\[
V_{mn}^{'}=
  \begin{bmatrix}
 \mathbb{V}_1  & \cdots  & \mathbb{V}_l
\end{bmatrix}
 \begin{bmatrix}
s^{\mathbb{V}\mathbb{I}}_{1,mn}\\ 
\vdots \\ 
s^{\mathbb{V}\mathbb{I}}_{l,mn}
\end{bmatrix}.
\]  
Thus $\Delta P_V=I_{mn}V_{mn}^{'}$.
$\hfill{} \Box$

\begin{corollary}
The change of total $I^2R$ loss, $\Delta P_I$, resulting from the parallel removal of a resistor edge with endpoint pair $\{m,n\}$ from a current-controlled circuit is given by $\Delta P_I=-I_{mn}V_{mn}^{'}$, where $I_{mn}$ denotes the current flowing on the edge before its removal, and $V_{mn}^{'}$ denotes the voltage difference between node pair $\{m,n\}$ after its removal. \label{prop:current}
\end{corollary} 
\textit{Proof:} At the end of the proof for Proposition \ref{prop:mix2}, we know that  $\Delta P_I$, the change of $I^2R$ loss  resulting from removing  $R_j$ from the current-controlled sub-circuit, is given by 
\[
\Delta P_I=
-I_{mn}\begin{bmatrix}
 \mathbb{I}_1  & \cdots  & \mathbb{I}_k
\end{bmatrix} 
 \begin{bmatrix}
s^{\mathbb{I}\mathbb{I}}_{1,mn}\\ 
\vdots \\ 
s^{\mathbb{I}\mathbb{I}}_{k,mn}
\end{bmatrix}.
\] 
Since there are no voltage sources  in a current-controlled circuit, we have
\[
V_{mn}^{'}=
 \begin{bmatrix}
 \mathbb{I}_1  & \cdots  & \mathbb{I}_k
\end{bmatrix} 
 \begin{bmatrix}
s^{\mathbb{I}\mathbb{I}}_{1,mn}\\ 
\vdots \\ 
s^{\mathbb{I}\mathbb{I}}_{k,mn}
\end{bmatrix}.
\] 
Thus $\Delta P_I=-I_{mn}V_{mn}^{'}$.
$\hfill{} \Box$

\begin{corollary}
The change of total $I^2R$ loss, $\Delta P_V$, resulting from the serial removal of a resistor edge with endpoint pair $\{m,n\}$ from a voltage-controlled circuit is given by $\Delta P_V=-I_{mn}^{'}V_{mn}$, where $V_{mn}$ denotes the voltage difference between node pair $\{m,n\}$ before its removal, and $I_{mn}^{'}$ denotes the current flowing through the short circuit edge connecting the node pair $\{m,n\}$ after the removal. \label{prop:voltage:serial}
\end{corollary} 
\textit{Proof:}  Since we are doing a  serial removal, it will merge node $m$ and node $n$ together. Electrically, it is  equivalent to  the parallel attachment of a zero resistance edge to $\{m,n\}$. Corollary \ref{prop:voltage} shows that the change of total $I^2R$ loss, $\Delta P_V$, resulting from the  parallel removal of a zero resistance edge from $\{m,n\}$ is $I_{mn}V_{mn}^{'}$.  In other words, it means the parallel attachment of a zero resistance edge to $\{m,n\}$ will change the total loss by  $-I_{mn}^{'}V_{mn}$. 
$\hfill{} \Box$

\begin{corollary}
The change of total $I^2R$ loss, $\Delta P_I$, resulting from the serial removal of a resistor edge with endpoint pair $\{m,n\}$ from a current-controlled circuit is given by $\Delta P_I=I_{mn}^{'}V_{mn}$, where $V_{mn}$ denotes the voltage difference between node pair $\{m,n\}$ before its removal, and $I_{mn}^{'}$ denotes the current flowing through the short circuit edge connecting the node pair $\{m,n\}$ after the removal. \label{prop:current:serial}
\end{corollary} 
\textit{Proof:}  Again, the serial removal of a resistor edge from $\{m,n\}$ is  equivalent to  the parallel attachment of a zero resistance edge to $\{m,n\}$. Corollary \ref{prop:current} shows that the change of total $I^2R$ loss, $\Delta P_I$, resulting from the  parallel removal of a zero resistance edge from $\{m,n\}$ is $-I_{mn}V_{mn}^{'}$.  In other words, it means the parallel attachment of a zero resistance edge to $\{m,n\}$ will change the total loss by  $I_{mn}^{'}V_{mn}$. 
$\hfill{} \Box$

\begin{corollary}
The change of total loss, $\Delta P_I$, resulting from adding a resistance edge  to a current-controlled network is equivalent to the change of total loss resulting from adding the resistance edge to its associated Norton equivalent circuit. The change of total loss, $\Delta P_V$, resulting from adding a resistance edge  to  a voltage-controlled network is equivalent to the change of total loss resulting from adding the resistance edge to its associated  Thevenin equivalent circuit.\label{prop:equcircuit}
\end{corollary} 
\textit{Proof:}  Based on Corollary \ref{prop:voltage} through Corollary \ref{prop:current:serial}, we know that in a single-source network the change of total loss resulting from the attachment (removal) of a resistor edge is completely determined by 
\begin{itemize}
\item the current flowing  through the link after (before) the change, and
\item the voltage difference between the endpoints of the edge before (after) the change.
\end{itemize}
In other words, the change of total loss is independent of the changes of  edges other than the one to be removed or added. This good property enables us to use the Thevenin equivalent circuit or the  Norton equivalent circuit to predict the change of total loss in a single-source circuit. Since we know the attachment of a link has  opposite effect on the loss of a current-controlled circuit and a voltage-controlled circuit, we'd better use Norton equivalent circuit to replace the current-controlled network and use Thevenin equivalent circuit to replace the voltage-controlled network. Of course, using Thevenin equivalent circuit to replace a current-controlled network is theoretically acceptable, but it requires an additional step to get the right answer, i.e. flipping the sign of the change of total loss. So is  using Norton equivalent circuit to replace a voltage-controlled network.
$\hfill{} \Box$

\begin{figure}[htbp]
	\centering
		\includegraphics[width=0.2\textwidth]{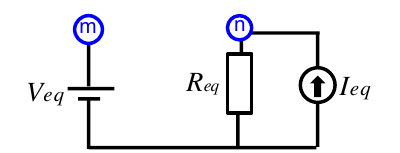}
	\caption{The equivalent circuit for a mixed-source network.} \label{equ:mix}
\end{figure}

\begin{corollary}
The change of total loss, $\Delta P$, resulting from adding a resistance edge  to  node pair $\{m,n\}$ in a mixed-source network is equivalent to the change of total loss, $\Delta P_{eq}$, resulting from adding the resistance edge to the equivalent circuit at terminal $m$-$n$ in Fig. 
\ref{equ:mix} where $I_{eq}$ is the current source in the Norton equivalent circuit of the current-controlled sub-circuit, $V_{eq}$ is the voltage source in the Thevenin equivalent circuit of the voltage-controlled sub-circuit, and $R_{eq}$ is the equivalent resistance. \label{prop:equcircuit:mix}
\end{corollary} 
\textit{Proof:}  Corollary \ref{prop:equcircuit} tells that we can use Norton equivalent circuit to replace the current-controlled sub-circuit and use Thevenin equivalent circuit to replace the voltage-controlled sub-circuit when calculating the change of total loss in each sub-circuit for a mixed-source one. Moreover, Proposition \ref{prop:mix2} and \ref{prop:serial} show that the change of total loss in a mixed-source circuit can be completely decomposed into two parts, one for its  voltage-controlled sub-circuit and one for its  current-controlled sub-circuit  \ref{prop:current:serial}. Thus, we just need to ``merge" together the Norton equivalent circuit of the  current-controlled sub-circuit and Thevenin equivalent circuit of the  voltage-controlled sub-circuit for the purposed of calculating the change of total loss for a mixed-source circuit. The idea is visualized in Fig. \ref{equ:mix}. 
$\hfill{} \Box$

\begin{remark}
Results similar to Corollary  \ref{prop:voltage} through  Corollary \ref{prop:current:serial}
hold for the removal of multiple resistors together, although
the proof becomes more involved.
\end{remark} 
\begin{remark}
A resistor in the circuit is always a passive  element as it consumes power. So $I_{mn}$ and $V_{mn}$ are always of opposite polarity. By Lemma \ref{lemma:polarity}, we know $V_{mn}$ and $V_{mn}^{'}$ are always of same polarity. Thus by  Corollary  \ref{prop:voltage} and  Corollary \ref{prop:current}, we know the parallel removal of a resistor edge  from a voltage-controlled circuit will always decrease total $I^2R$ loss, and the parallel removal of a resistor edge  from a current-controlled circuit will always increase total $I^2R$ loss. It is worth mentioning that this is much deeper than the formulas  for serial ($R=R_1+R_2$)  and parallel ($R=\frac{1}{1/R_1+1/R_2}$) connection of resistors,  as the removal of a resistor not only change the resistance of the graph but also redistribute the currents in the graph. Both the above results are consistent with \cite{Baillieul} and \cite{Wang}.
\end{remark} 
\begin{proposition} 
The change of total losses, $\Delta P$, resulting from the parallel removal (parallel attachment) of a current source $\mathbb{I}_j$  from  (to)  a mixed-source circuit is given by $\Delta P= \Delta P_I$, where  $\Delta P_I$ denotes the change of losses resulting from removing  (adding) the link  $\mathbb{I}_j$ from its current-controlled sub-circuit.\label{prop:source}
\end{proposition} 
\textit{Proof:} As in Proposition \ref{prop:mix2}, we just need to prove the parallel removal part of the proposition.

Again, we assume   there are $k$ current sources $\{\mathbb{I}_1,...,\mathbb{I}_k\}$, and $l$ voltage sources $\{\mathbb{V}_1,...,\mathbb{V}_l\}$, in the circuit. Suppose we are going to remove the last current source $\mathbb{I}_k$, and its endpoint pair is $\{m,n\}$. 

By the principle of energy conservation, the change of total $I^2R$ loss must be equivalent to the change of total sources' energy output  
\[\sum_{i=1}^{k-1}\mathbb{I}_i\Delta V_i-\mathbb{I}_k  V_{k}+\sum_{i=1}^{l}\mathbb{V}_i\Delta I_i
\]
where $\Delta V_i$ denotes the change of voltage difference between the endpoints of the $i$-th current source, $V_{k}$ denotes the voltage difference between the endpoints of the $k$-th current source before the removal, and $\Delta I_i$ denotes the change of current flowing through   the $i$-th voltage source. 

By the definition of resistance graph, we know putting back all  sources will just create some short circuits and open circuits which essentially don't change the resistance graph. In addition, the sensitivity of the current flowing through an edge (or of voltage difference between the endpoints of an edge) is completely determined by the resistance graph, i.e. constant resistance graph means constant source factors. Thus by the superposition principle, we know $V_{k}$ is a linear combination of $\{\mathbb{I}_1,...,\mathbb{I}_k\}$ and $\{\mathbb{V}_1,...,\mathbb{V}_l\}$, i.e.
\[
V_{k}=
\begin{bmatrix}
 \mathbb{I}_1  & \cdots  & \mathbb{I}_k  
\end{bmatrix}\begin{bmatrix}
s^{\mathbb{I}\mathbb{I}}_{1,k}\\ 
\vdots \\ 
 s^{\mathbb{I}\mathbb{I}}_{k,k}
\end{bmatrix}
+\begin{bmatrix}
 \mathbb{V}_1  & \cdots  & \mathbb{V}_l  
\end{bmatrix}\begin{bmatrix}
s^{\mathbb{V}\mathbb{I}}_{1,k}\\ 
\vdots \\ 
 s^{\mathbb{V}\mathbb{I}}_{l,k}\end{bmatrix}
\]
where $\{s^{\mathbb{I}\mathbb{I}}_{1,k} , \cdots , s^{\mathbb{I}\mathbb{I}}_{k,k}\} $ and $\{s^{\mathbb{V}\mathbb{I}}_{1,k} , \cdots , s^{\mathbb{V}\mathbb{I}}_{l,k}\} $ are source factors. Following a procedure  similar to  Proposition \ref{prop:mix2}, we have  
\[
\begin{bmatrix}
\Delta V_1\\ 
\vdots \\ 
\Delta V_{k-1}
\end{bmatrix}=
-\mathbb{I}_{k}\begin{bmatrix}
s^{\mathbb{I}\mathbb{I}}_{k,1}\\ 
\vdots \\ 
s^{\mathbb{I}\mathbb{I}}_{k,k-1}
\end{bmatrix}\]\[
\begin{bmatrix}
\Delta I_1\\ 
\vdots \\ 
\Delta I_l
\end{bmatrix}=
-\mathbb{I}_{k}\begin{bmatrix}
s^{\mathbb{I}\mathbb{V}}_{k,1}\\ 
\vdots \\ 
s^{\mathbb{I}\mathbb{V}}_{k,l}
\end{bmatrix}.
\] 
By (\ref{equ:factor}), we have  $s^{\mathbb{I}\mathbb{I}}_{k,i}=s^{\mathbb{I}\mathbb{I}}_{i,k}$ ($i=1,...,k$) and   $s^{\mathbb{I}\mathbb{V}}_{k,i}=-s^{\mathbb{V}\mathbb{I}}_{i,k}$ ($i=1,...,l$)
which   gives us 
\[
\begin{split}
&\sum_{i=1}^{k-1}\mathbb{I}_i\Delta V_i-\mathbb{I}_k  V_{k}+\sum_{i=1}^{l}\mathbb{V}_i\Delta I_i=
-\mathbb{I}_{k}\{\begin{bmatrix}
 \mathbb{I}_1 \ \cdots \ \mathbb{I}_{k-1}
\end{bmatrix} 
 \begin{bmatrix}
s^{\mathbb{I}\mathbb{I}}_{1,k}\\ 
\vdots \\ 
s^{\mathbb{I}\mathbb{I}}_{k-1,k}
\end{bmatrix}
+\begin{bmatrix}
 \mathbb{I}_1 \ \cdots  \ \mathbb{I}_k  
\end{bmatrix}\begin{bmatrix}
s^{\mathbb{I}\mathbb{I}}_{1,k}\\ 
\vdots \\ 
 s^{\mathbb{I}\mathbb{I}}_{k,k}
\end{bmatrix}\}.
\end{split}
\]

As we can see, the above result is independent of the voltage sources in the mixed-source circuit. By setting the value of all voltage sources to zero, we get the current-controlled sub-circuit. It is easy to prove that  $\Delta P_I$, the change of $I^2R$ loss  resulting from removing  $\mathbb{I}_k$ from the current-controlled sub-circuit, is   equivalent to above result.
$\hfill{} \Box$

Using the same idea, we can get a similar proposition for   the serial removal (serial attachment) of a voltage source   from a mixed-source circuit.
\begin{proposition} 
The change of total losses, $\Delta P$, resulting from the serial removal (serial attachment) of a voltage source $\mathbb{V}_j$  from (to) a mixed-source circuit is given by $\Delta P= \Delta P_V$, where  $\Delta P_V$ denotes the change of losses resulting from removing  (adding) the link  $\mathbb{V}_j$ from its voltage-controlled sub-circuit. \label{removevoltagemix}
\end{proposition} 
\textit{Proof:} As in Proposition \ref{prop:mix2}, we just need to prove the parallel removal part of the proposition.

Again, we assume   there are $k$ current sources $\{\mathbb{I}_1,...,\mathbb{I}_k\}$, and $l$ voltage sources $\{\mathbb{V}_1,...,\mathbb{V}_l\}$, in the circuit. Suppose we are going to remove the last current source $\mathbb{V}_l$, and its endpoint pair is $\{m,n\}$. 

By the principle of energy conservation, the change of total $I^2R$ loss must be equivalent to the change of total sources' energy output  
\[\sum_{i=1}^{k}\mathbb{I}_i\Delta V_i+\sum_{i=1}^{l-1}\mathbb{V}_i\Delta I_i-\mathbb{V}_l  I_{l}
\]
where $\Delta V_i$ denotes the change of voltage difference between the endpoints of the $i$-th current source, $I_{l}$ denotes the current flowing through the $l$-th voltage source before the removal, and $\Delta I_i$ denotes the change of current flowing through   the $i$-th voltage source. 

Again, by the superposition principle, we know $I_{l}$ is a linear combination of $\{\mathbb{I}_1,...,\mathbb{I}_k\}$ and $\{\mathbb{V}_1,...,\mathbb{V}_l\}$, i.e.
\[
I_{l}=
\begin{bmatrix}
 \mathbb{I}_1  & \cdots  & \mathbb{I}_k  
\end{bmatrix}\begin{bmatrix}
s^{\mathbb{I}\mathbb{V}}_{1,l}\\ 
\vdots \\ 
 s^{\mathbb{I}\mathbb{V}}_{k,l}
\end{bmatrix}
+\begin{bmatrix}
 \mathbb{V}_1  & \cdots  & \mathbb{V}_l  
\end{bmatrix}\begin{bmatrix}
s^{\mathbb{V}\mathbb{V}}_{1,l}\\ 
\vdots \\ 
 s^{\mathbb{V}\mathbb{V}}_{l,l}\end{bmatrix}
\]
where $\{s^{\mathbb{I}\mathbb{V}}_{1,l} , \cdots , s^{\mathbb{I}\mathbb{V}}_{k,l}\} $ and $\{s^{\mathbb{V}\mathbb{V}}_{1,l} , \cdots , s^{\mathbb{V}\mathbb{V}}_{l,l}\} $ are source factors. Following a procedure  similar to  Proposition \ref{prop:mix2}, we have  
\[
\begin{bmatrix}
\Delta V_1\\ 
\vdots \\ 
\Delta V_{k}
\end{bmatrix}=
-\mathbb{V}_l\begin{bmatrix}
s^{\mathbb{V}\mathbb{I}}_{l,1}\\ 
\vdots \\ 
s^{\mathbb{V}\mathbb{I}}_{l,k}
\end{bmatrix}
\]
\[
\begin{bmatrix}
\Delta I_1\\ 
\vdots \\ 
\Delta I_{l-1}
\end{bmatrix}=
-\mathbb{V}_l\begin{bmatrix}
s^{\mathbb{V}\mathbb{V}}_{l,1}\\ 
\vdots \\ 
s^{\mathbb{V}\mathbb{V}}_{l,l-1}
\end{bmatrix}.
\] 
By (\ref{equ:factor}), we have  $s^{\mathbb{V}\mathbb{V}}_{l,i}=s^{\mathbb{V}\mathbb{V}}_{i,l}$ ($i=1,...,l$) and   $s^{\mathbb{V}\mathbb{I}}_{l,i}=-s^{\mathbb{I}\mathbb{V}}_{i,l}$ ($i=1,...,k$)
which   gives us 
\[
\begin{split}
&\sum_{i=1}^{k}\mathbb{I}_i\Delta V_i+\sum_{i=1}^{l-1}\mathbb{V}_i\Delta I_i-\mathbb{V}_l  I_{l}= 
-\mathbb{V}_l \begin{bmatrix}
 \mathbb{V}_1  & \cdots  & \mathbb{V}_{l-1}
\end{bmatrix} 
 \begin{bmatrix}
s^{\mathbb{V}\mathbb{V}}_{1,l}\\ 
\vdots \\ 
s^{\mathbb{V}\mathbb{V}}_{l-1,l}
\end{bmatrix}
-\mathbb{V}_l \begin{bmatrix}
 \mathbb{V}_1  & \cdots  & \mathbb{V}_l  
\end{bmatrix}\begin{bmatrix}
s^{\mathbb{V}\mathbb{V}}_{1,l}\\ 
\vdots \\ 
 s^{\mathbb{V}\mathbb{V}}_{l,l}
\end{bmatrix}.
\end{split}
\]

As we can see, the above result is independent of the current sources in the mixed-source circuit. By setting the value of all current sources to zero, we get the voltage-controlled sub-circuit. It is easy to prove that  $\Delta P_V$, the change of $I^2R$ loss  resulting from removing  $\mathbb{V}_l$ from the voltage-controlled sub-circuit, is   equivalent to above result.
$\hfill{} \Box$

By the definition of resistance graph, we know a mixed-source network can be created by the parallel attachment of all current sources and the serial attachment of all voltages to the resistance graph. Thus, combining the result of Proposition \ref{prop:source} and Proposition \ref{removevoltagemix}, we have the following:
\begin{proposition} (Superposition of $I^2R$ loss)
The  total $I^2R$ loss, $P$, of a mixed-source circuit is given by $ P =P_I+P_V$, where $P_I$ denotes the total $I^2R$ loss of its current-controlled sub-circuit, and $P_V$ denotes  the total $I^2R$ loss of its voltage-controlled sub-circuit. \label{prop:sumloss}
\end{proposition} 

\begin{corollary}
The change of total $I^2R$ loss, $\Delta P_I$, resulting from the parallel removal of a current source $\mathbb{I}$ with endpoint pair $\{m,n\}$ from a current-controlled circuit  is given by $\Delta P_I=-\mathbb{I}(V_{m,n}+V_{m,n}^{'})$, where $V_{m,n}$ and $V_{m,n}^{'}$ denote the voltage differences between node pair $\{m,n\}$ before and after its removal, respectively. \label{removecurrent}
\end{corollary} 
\textit{Proof:} Again, we assume   there are $k$ current sources $\{\mathbb{I}_1,...,\mathbb{I}_k\}$ in the circuit. Suppose we are going to remove the last current source $\mathbb{I}_k$, and its endpoint pair is $\{m,n\}$. At the end of the proof of Proposition  \ref{prop:source}, we know
\[
\begin{split}
&\Delta P_I= 
-\mathbb{I}_{k}\begin{bmatrix}
 \mathbb{I}_1  & \cdots  & \mathbb{I}_{k-1}
\end{bmatrix} 
 \begin{bmatrix}
s^{\mathbb{I}\mathbb{I}}_{1,k}\\ 
\vdots \\ 
s^{\mathbb{I}\mathbb{I}}_{k-1,k}
\end{bmatrix}
-\mathbb{I}_{k} \begin{bmatrix}
 \mathbb{I}_1  & \cdots  & \mathbb{I}_k  
\end{bmatrix}\begin{bmatrix}
s^{\mathbb{I}\mathbb{I}}_{1,k}\\ 
\vdots \\ 
 s^{\mathbb{I}\mathbb{I}}_{k,k}
\end{bmatrix}.
\end{split}
\]
Also we know 
\[
V_{m,n}=
 \begin{bmatrix}
 \mathbb{I}_1  & \cdots  & \mathbb{I}_{k}
\end{bmatrix} 
 \begin{bmatrix}
s^{\mathbb{I}\mathbb{I}}_{1,k}\\ 
\vdots \\ 
s^{\mathbb{I}\mathbb{I}}_{k,k}
\end{bmatrix}
\]
and
\[
V_{m,n}^{'}=
 \begin{bmatrix}
 \mathbb{I}_1  & \cdots  & \mathbb{I}_{k-1}
\end{bmatrix} 
 \begin{bmatrix}
s^{\mathbb{I}\mathbb{I}}_{1,k}\\ 
\vdots \\ 
s^{\mathbb{I}\mathbb{I}}_{k-1,k}
\end{bmatrix}.
\]
Thus
$\Delta P_I=-\mathbb{I}(V_{m,n}+V_{m,n}^{'})$.
$\hfill{} \Box$
\begin{corollary}
The change of total $I^2R$ loss, $\Delta P_V$, resulting from the serial removal of a voltage source $\mathbb{V}$ with endpoint pair $\{m,n\}$ from a voltage-controlled circuit  is given by $\Delta P_V=-\mathbb{V}(I_{m,n}+I_{m,n}^{'})$, where $I_{m,n}$ denotes the current that flowed through the voltage source before the removal, and $I_{mn}^{'}$ denotes the current flowing through the short circuit edge connecting the node pair $\{m,n\}$ after the removal.\label{removevoltage}
\end{corollary} 
\textit{Proof:} Again, we assume   there are   $l$ voltage sources $\{\mathbb{V}_1,...,\mathbb{V}_l\}$, in the circuit. Suppose we are going to remove the last current source $\mathbb{V}_l$, and its endpoint pair is $\{m,n\}$.  At the end of the proof of Proposition  \ref{removevoltagemix}, we know
\[
\begin{split}
&\Delta P_V=
-\mathbb{V}_l \begin{bmatrix}
 \mathbb{V}_1  & \cdots  & \mathbb{V}_{l-1}
\end{bmatrix} 
 \begin{bmatrix}
s^{\mathbb{V}\mathbb{V}}_{1,l}\\ 
\vdots \\ 
s^{\mathbb{V}\mathbb{V}}_{l-1,l}
\end{bmatrix}
-\mathbb{V}_l \begin{bmatrix}
 \mathbb{V}_1  & \cdots  & \mathbb{V}_l  
\end{bmatrix}\begin{bmatrix}
s^{\mathbb{V}\mathbb{V}}_{1,l}\\ 
\vdots \\ 
 s^{\mathbb{V}\mathbb{V}}_{l,l}
\end{bmatrix}.
\end{split}
\]
Also we know
\[
I_{mn}= 
 \begin{bmatrix}
 \mathbb{V}_1  & \cdots  & \mathbb{V}_l  
\end{bmatrix}\begin{bmatrix}
s^{\mathbb{V}\mathbb{V}}_{1,l}\\ 
\vdots \\ 
 s^{\mathbb{V}\mathbb{V}}_{l,l}
\end{bmatrix} 
\]
and
\[
I_{mn}^{'}= 
 \begin{bmatrix}
 \mathbb{V}_1  & \cdots  & \mathbb{V}_{l-1}  
\end{bmatrix}\begin{bmatrix}
s^{\mathbb{V}\mathbb{V}}_{1,l}\\ 
\vdots \\ 
 s^{\mathbb{V}\mathbb{V}}_{l-1,l}
\end{bmatrix} .
\]
Thus $\Delta P_V=-\mathbb{V}(I_{m,n}+I_{m,n}^{'})$.
$\hfill{} \Box$

\begin{remark}
Results similar  to Corollary  \ref{removecurrent} through  Corollary \ref{removevoltage}
hold for the removal of multiple sources together, although
the proof becomes more involved.
\end{remark} 
 \begin{remark}
Proposition \ref{prop:source} and Corollary  \ref{removecurrent} together explain the reason for the paradoxical behavior in Fig. \ref{fig:paradox}. After we create the current-controlled  sub-circuit of Fig. \ref{fig:paradox}(a), we can compute the voltage difference between the node pair from which the 0.25A current source is removed. The voltage difference between this node pair is 0V and 0.125V before and after the removal, respectively.  So the total loss will decrease by  $0.25(0+0.125)= 0.03125$W.
\end{remark}

\begin{remark} (Orthogonality between the effects of voltage sources and current sources) Suppose there are $n$ resistance edges  $\{R_1,...,R_n\}$ in the mixed-source circuit, and we denote the currents flowing on the $i$-th resistance edge as $I_i^{M}$ in the mixed-source circuit, $I_i^{V}$ in its voltage-controlled sub-circuit, and $I_i^{I}$ in its current-controlled sub-circuit, respectively. The diagonal matrix of edge resistances is 
\[\Lambda =diag\{R_1,...,R_n\}
\]
Write the vector of currents flowing through each edge as $\vec{I}_M=[I_1^{M},...,I_n^{M}]^{T}$ in the mixed-source circuit,   $\vec{I}_V=[I_1^{V},...,I_n^{V}]^{T}$ in the voltage-controlled sub-circuit, and $\vec{I}_I=[I_1^{I},...,I_n^{I}]^{I}$ in the voltage-controlled sub-circuit, respectively. 
Then by the superposition principle, we must have
\[
\begin{split}
{\vec{I}_M}^{T} \Lambda \vec{I}_M &=({\vec{I}_V}^{T} +{\vec{I}_I}^{T} )\Lambda ({\vec{I}_V}  +{\vec{I}_I}  )\\&
=  {\vec{I}_V}^{T} \Lambda  {\vec{I}_V}  +{\vec{I}_I}^{T} \Lambda  {\vec{I}_I}+2{\vec{I}_V}^{T} \Lambda  {\vec{I}_I} .
\end{split}
\]
Combining the above result with Proposition \ref{prop:sumloss}, we must have
\[
{\vec{I}_V}^{T} \Lambda  {\vec{I}_I} =0.
\]
\end{remark}

\section{Four Equivalent Loss Computing Methods}
Based on the discussion in this paper, we   have four different methods to calculate the total $I^2R$ loss of an arbitrary mixed-source circuit. To describe and compare them, we assume  there are $k$ current sources $\{\mathbb{I}_1,...,\mathbb{I}_k\}$, $l$ voltage sources $\{\mathbb{V}_1,...,\mathbb{V}_l\}$, and $t$ resistors $\{\mathbb{R}_1,...,\mathbb{R}_t\}$ in the circuit.
\begin{itemize}
\item[(a)] The first way is the most traditional one. We calculate either  the total power output of all voltage sources and current sources or the total $I^2R$ loss of all resistors which is given by
\begin{equation}
\begin{split}
&\sum_{i=1}^{k}\mathbb{I}_i V_i +\sum_{i=1}^{l}\mathbb{V}_i I_i\ \ \ (or\ \sum_{i=1}^{t}P_{\mathbb{R}_i}) 
\\ s.t. \ \ & Kirchhoff \ voltage\  laws
\\& Kirchhoff \ current\  laws
\end{split}
\end{equation}
where $V_{i}$ denotes the voltage difference between the endpoints of the $i$-th current source before the removal,  $\ I_i$ denotes the  current flowing through   the $i$-th voltage source before the removal, and $P_{\mathbb{R}_i}$ denotes the $I^2R$ loss of the $i$-th resistor.  $\mathbb{I}_i$ and $\mathbb{V}_i$ are the controlling currents and voltages respectively, they remain constant under the considered topology reconfiguration. 
\item[(b)]  The second way is based on the following idea: a mixed-source circuit can be created by first constructing the current-controlled sub-circuit and then putting back all voltage sources. In \cite{Wang}, we show that the partial derivative of the cost function $P_I$  described by (\ref{loss:current})  in the direction $I_{e_i}$ is exactly double the algebraic sum of all voltages on the fundamental cycle defined by $e_i$. 
This means we are including the constraints associated with Kirchhoff's voltage law and the voltage sources $\mathbb{V}_i$ relative to fundamental cycles in which they appear. The operating point is thus given by minimizing cost function $P_I$ with   linear constraints denoting the effect of voltage sources.
\begin{equation}
\begin{split}
&min \ P_I
\\ s.t. \ \ & some\ linear\ constraints\ denoting\ the\ effect\ of\ voltage\  sources\ are\ satisfied
\end{split}
\end{equation}
\item[(c)]  The third way is quite similar to the second way: a mixed-source circuit can be created by first constructing the voltage-controlled sub-circuit and then putting back all current sources. Following an idea similar to (b), the total loss  can  be given by minimizing cost function $P_V$ described in (\ref{loss:voltage}) with   linear constraints denoting the effect of current sources.
\begin{equation}
\begin{split}
&min \ P_V
\\ s.t. \ \ & some\ linear\ constraints\ denoting\ the\  effect\ of\ current\ sources\ are\ satisfied
\end{split}
\end{equation}
\item[(d)]  The fourth way is based on Proposition \ref{prop:sumloss} and is given by
\begin{equation}
min \  P_V+ min \ P_I
\end{equation}
where $P_V$ is the potential function (described by (\ref{loss:voltage})) of the voltage-controlled sub-circuit, and  $P_I$ is the potential function (described by (\ref{loss:current})) of the current-controlled sub-circuit. 
\end{itemize}

Although the four methods are of different mathematical  forms, they must give us the same result. Specifically, methods (b), (c), and (d) are closely related to each other. In mixed source networks, the operating values of voltages and currents are determined as critical values of $P_V$ (as a quadratic function of fundamental node variables) and $P_I$ (as a quadratic function of fundamental cycle variables) where it is assumed that all voltage sources $\mathbb{V}_i$ and current sources $\mathbb{I}_i$ are present. This is the approach of methods (b) and (c). This approach can be carried out by solving for critical points of $P_I$ (with respect to the fundamental cycles variables) subject to the Kirchhoff voltage constraints  that are obtained by adding the $\mathbb{V}_i$ to the resistance graph. Similarly, one can solve for the  critical points of $P_V$ (with respect to the fundamental node variables) subject to the  Kirchhoff current constraints  by including the  $\mathbb{I}_i$ to the resistance graph. Method (d) utilizes the novel decomposition of mixed-source circuit in Proposition \ref{prop:sumloss} and integrates  (b) and (c) together. By calculating the loss of the voltage-controlled sub-circuit $P_V$ and the loss of the current-controlled sub-circuit $P_I$ individually, method (d) solves for the loss of a mixed-source circuit in an unconstrained quadratic programming  form.

\section{Conclusion}
Our previous work, \cite{Baillieul} and \cite{Wang}, proved that in a single source network, although the detailed change of loss is impossible to predict  without solving the Kirchhoff's equations, the sign of the overall change in  $I^2R$  loss is always certain. Such predictability with respect to total loss is not present in a mixed-source network, and this may result in interesting paradoxes such as fewer sources producing more power. While the demonstrated uncertainty to changes in a mixed-source circuit together with the well recognized complexity in line switching suggest that active control of grid topology in a mixed-source model is a formidable problem, our results nevertheless offer a clean decomposition for the mixed-source circuit that completely separates the effect of current sources and voltage sources on network total loss. As the world's power grids increasingly embrace novel energy sources and new classes of assets associated with storage and demand response, our ongoing research seeks to use the decomposition concepts we have presented for developing new approaches to resource allocation that appropriately balance generation scheduling, grid topology configuration, and recruitment of demand response.
\bibliography{references}
\bibliographystyle{IEEEtran}
 
\end{document}